\documentclass[cameraready]{Interspeech}

\title{GenTSE: Enhancing Target Speaker Extraction via a Coarse-to-Fine Generative Language Model}

\author[affiliation={1}]{Haoyang}{Li}
\author[affiliation={1}]{Xuyi}{Zhuang}
\author[affiliation={1}]{Azmat}{Adnan}
\author[affiliation={2}]{Ye}{Ni}
\author[affiliation={1}]{Wei}{Rao}
\author[affiliation={1}]{Shreyas}{Gopal}
\author[affiliation={1}]{Eng Siong}{Chng}
\author[affiliation={3,4}]{Boon Siew}{Han}
\author[affiliation={1}]{Yuanjin}{Zheng}

\address{
    $^1$ Nanyang Technological University, Singapore \\
    $^2$ Southeast University, China \\
    $^3$ Schaeffler Hub for Advanced REsearch (SHARE) at Nanyang Technological University, Singapore \\
    $^4$ Schaeffler
}

\email{li0078ng@e.ntu.edu.sg}


\usepackage{comment}
\usepackage{graphicx}
\usepackage{multirow}
\usepackage{makecell} 
\usepackage{cite}
\usepackage{adjustbox}
\usepackage{booktabs}
\usepackage{array}
\usepackage{caption}
\usepackage{subcaption}  
\usepackage{tabularx}
\usepackage{amsmath}
\usepackage{color}
\usepackage{amssymb}
\usepackage{hyperref}
\usepackage{xurl}

\begin{document}

\maketitle

\begin{abstract}
Language Model (LM)-based generative modeling has emerged as a promising direction for TSE, offering potential for improved generalization and high-fidelity speech. We propose GenTSE, a two-stage decoder-only generative LM for TSE: Stage-1 predicts coarse semantic tokens, and Stage-2 generates fine acoustic tokens. Separating semantics and acoustics stabilizes decoding and yields more accurate target speech. Both stages use continuous SSL or codec embeddings, offering richer context than discretized-prompt methods. To reduce exposure bias, we employ a Frozen-LM Conditioning training strategy that conditions the LMs on predicted tokens from earlier checkpoints to reduce the gap between teacher-forcing training and autoregressive inference. We further apply DPO to better align outputs with perceptual preferences. Experiments on Libri2Mix show that GenTSE surpasses previous LM-based systems in speech quality, intelligibility, and speaker consistency.

\end{abstract}


\section{Introduction}
\label{sec:intro}
Target Speaker Extraction (TSE) aims to recover a designated speaker’s voice from a multi-speaker mixture using auxiliary information that identifies the target speaker \cite{zmolikova2023neural}. Most existing TSE methods rely on discriminative, supervised models that learn a direct mapping from the mixture to the target speech \cite{xu2020spex, ge2020spex+, chen2023mc, he2024hierarchical, zeng2025usef}, but such approaches generalize poorly under distribution shifts and compromise the fidelity of the extracted speech signal. Generative models such as Diffusion and Language Model, which learn the distribution of a target signal conditioned on relevant contextual information, have emerged as promising alternatives due to their strong capability to generalize to unseen data \cite{wang2025solospeech} and synthesize speech with high fidelity \cite{zhang2025anyenhance}. 

Recent LM-based methods have explored generative formulations for TSE. LLaSE-G1 \cite{kang2025llase} and UniSE \cite{yan2025unise} adopt single-stage LMs to predict codec tokens, but direct modeling of fine acoustics is challenging because the LM must jointly learn content, speaker traits, and detailed signal structure within a high-entropy token space. AnyEnhance \cite{zhang2025anyenhance} uses a two-stage pipeline with a semantic stage based on continuous embedding regression, while TSELM \cite{tang2024tselm} applies a pre-LM cross-attention fusion block to a single-stage LM; in both cases, these components introduce non-generative components. LauraTSE \cite{LauraTSE} predicts coarse audio codec tokens autoregressively, but restores fine details with a deterministic encoder-only Transformer, similarly breaking end-to-end generativity. SpeechX \cite{wang2024speechx} and Metis \cite{wang2025metis} use discretized reference prompts, which lose fine-grained speaker information. Unlike prior works, we propose GenTSE, a fully generative two-stage decoder-only LM architecture: Stage-1 generates coarse semantic tokens, and Stage-2 predicts fine acoustic tokens conditioned on them. Both stages are further conditioned on continuous SSL or codec embeddings, forming a coherent coarse-to-fine generative hierarchy.

We further address exposure bias, an overlooked yet critical issue in autoregressive (AR) LM-based SE and TSE that degrades speech fidelity and speaker consistency. Exposure bias arises from the mismatch between teacher-forcing (TF) during training and autoregressive inference, where the model must rely on its own past predictions. Inspired by scheduled sampling in neural speech translation \cite{mihaylova2019scheduled, liu2021scheduled}, we optimize GenTSE with Frozen-LM Conditioning (FLC), which explicitly exposes the models to predicted tokens from earlier frozen checkpoints during training, thus narrowing the training-inference mismatch.

Finally, we apply preference alignment via Direct Preference Optimization (DPO) \cite{rafailov2023direct}. Existing TSE approaches \cite{liu2023x, zeng2025usef} typically optimize signal-based objectives (e.g. SI-SDR or likelihood), which reduce prediction error but do not reliably capture human-perceived naturalness, clarity, or speaker fidelity. Although \cite{itani2025neural} incorporates human feedback through a refinement network, it still depends on low-level signal proxies and introduces additional inference-time parameters. In contrast, DPO offers a simple and stable way to align model output with human preference without modifying the inference architecture. Inspired by its recent use in speech enhancement \cite{li2025aligning}, we adapt DPO to TSE for the first time and show that it effectively guides the model's output toward higher perceived quality. On Libri2Mix \cite{cosentino2020librimix}, GenTSE shows promising performance in speech quality, intelligibility, and speaker consistency. 


\begin{figure*}[!t]
    \centering
    \includegraphics[scale=0.10]{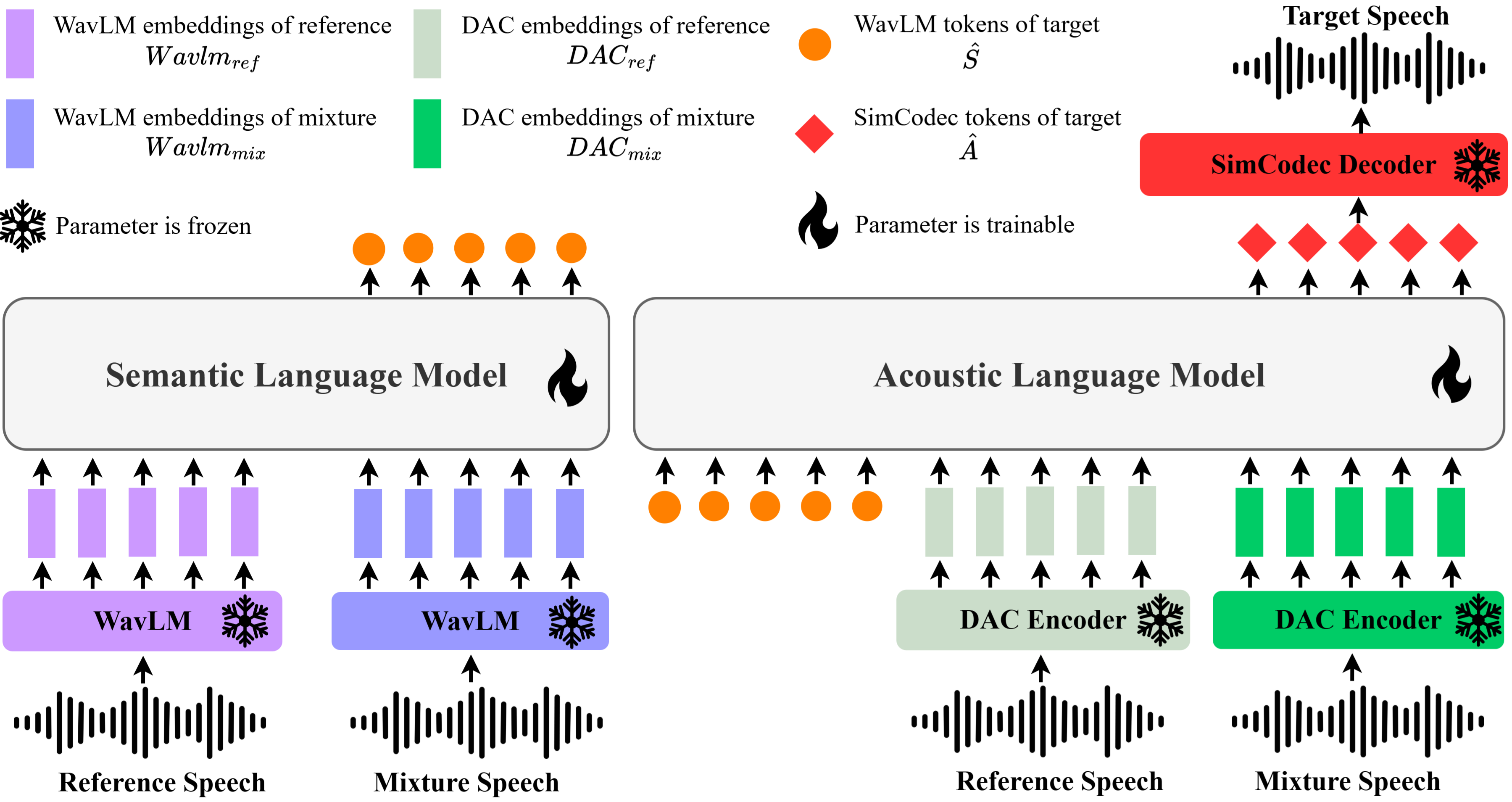}
    \caption{The overall architecture of GenTSE consisting of a semantic extraction stage (left) and an acoustic generation stage (right)}
    \label{fig:overall_pipeline} 
\end{figure*}

\section{Methodology}
\label{sec: method}
Fig. \ref{fig:overall_pipeline} shows the proposed GenTSE framework, which comprises a semantic extraction stage followed by an acoustic generation stage. The semantic stage autoregressively predicts coarse semantic tokens for the target speaker, and the acoustic stage generates fine-grained acoustic tokens conditioned on them. Trained independently, the two stages form a fully generative hierarchical pipeline that avoids directly modeling low-level acoustics, improving speaker extraction. Section~\ref{subsec: Hierarchical Modeling} details the overall framework, while Section~\ref{subsec: Frozen-LM Conditioning} and Section~\ref{subsec: Direct Preference Optimization} introduce two novel optimization techniques in TSE: Frozen-LM Conditioning (FLC) to reduce exposure bias and Direct Preference Optimization (DPO) to enhance perceptual quality. 

\subsection{Hierarchical Modeling}
\label{subsec: Hierarchical Modeling}

\subsubsection{Semantic extraction stage}
Given a reference and mixture speech, WavLM \cite{chen2022wavlm} serves as the high level feature extractor to extract the corresponding frame-level embeddings from layer 6, denoted as $Wavlm_{ref}$ , $Wavlm_{mix}\in \mathbb{R}^{T \times H}$, where $T$ and $H$ are the time and hidden dimensions. An AR decoder-only LM then conditions on $Wavlm_{ref}$ and $Wavlm_{mix}$ to generate $\hat{S}=[\hat{s}_1, \hat{s}_2, \ldots, \hat{s}_T]$, the predicted semantic tokens of the target speech signal. The corresponding ground-truth tokens $S$ are obtained from the target signal by first extracting frame-level embeddings from layer 6 of WavLM followed by discretization via kmeans. The baseline training objective of the LM employs cross-entropy (CE) loss with teacher-forcing following Eq. (\ref{eq:ce_loss_sem}).

\begin{equation}
\mathcal{L}_{\text{CE}_{\text{sem}}}
= - \sum_{t=1}^{T}
\log P_\theta(
    \hat{s}_t \mid s_{<t}, Wavlm_{ref}, Wavlm_{mix}).
\label{eq:ce_loss_sem}
\end{equation}

\subsubsection{Acoustic generation stage}
To recover low-level acoustic details, the DAC \cite{kumar2023high} encoder is used to extract rich acoustic features from reference and mixture speech signals, obtaining embeddings $DAC_{ref}$, $DAC_{mix}\in \mathbb{R}^{T \times H}$. The semantic tokens of the target speech ($S$ during training, $\hat{S}$ during inference) are concatenated with $DAC_{ref}$ and $DAC_{mix}$ as the condition for another AR decoder-only LM to generate $\hat{A}=[\hat{a}_1, \hat{a}_2, \ldots, \hat{a}_N]$, the predicted acoustic tokens of the target speech signal. We use SimCodec \cite{yao2025gense}, a neural audio codec with a single codebook as the acoustic tokenizer. Using a single-codebook codec simplifies the LM by eliminating the need to jointly model multiple token streams, unlike prior multi-layer codebook approaches \cite{zhang2025anyenhance, wang2025metis}. Finally, the Simcodec decoder reconstructs target speech signal from $\hat{A}$. The baseline training objective follows Eq. (\ref{eq:ce_loss_aco}).

\begin{equation}
\mathcal{L}_{\text{CE}_{\text{aco}}}
= - \sum_{n=1}^{N}
\log P_\phi(
    \hat{a}_n \mid a_{<n}, S, DAC_{ref}, DAC_{mix}).
\label{eq:ce_loss_aco}
\end{equation}

\begin{table*}[t]
  \caption{Libri2Mix (clean) results. “G”/“D” denote generative/discriminative models; * indicates models reproduced on our dataset.}
  \renewcommand{\arraystretch}{1.2}
  \begin{center}
    \resizebox{\textwidth}{!}{
    \begin{tabular}{ccccccccccc}
      \Xhline{2\arrayrulewidth}
      \multirow{2}{*}{Model} & \multirow{2}{*}{Category} & \multicolumn{3}{c}{DNSMOS $\uparrow$} & \multirow{2}{*}{UTMOS $\uparrow$} & \multirow{2}{*}{NISQA $\uparrow$} & \multirow{2}{*}{SECS $\uparrow$} & \multirow{2}{*}{dWER $\downarrow$} & \multirow{2}{*}{SpeechBERT $\uparrow$} \\
      \cline{3-5}
                             &                           & SIG & BAK & OVL &                         &                      &                         &                          &                          \\
      \hline
      Mixture & - & 3.383 & 3.098 & 2.653 & 1.519 & 2.251 & 0.754 & 0.821 & 0.655 \\
      \hline
      X-TF-GridNet\cite{hao2024x}  & D & 3.328 & 3.607 & 2.895 & 2.852 & 3.216 & 0.758 & 0.389 & 0.784 \\
      USEF-SepFormer* \cite{zeng2025usef} & D & 3.324 & 3.698 & 2.927 & 3.492 & 2.880 & 0.806 & \textbf{0.156} & 0.830 \\
      \hline
      TSELM-L* \cite{tang2024tselm} & G & 3.478 & 4.035 & 3.198 & 3.556 & 3.509 & 0.651 & 0.263 & 0.832 \\
      LLaSE-G1 \cite{kang2025llase} & G & 3.531 & 4.015 & 3.226 & 3.228 & 3.638 & 0.839 & 0.476 & 0.825 \\
      Metis \cite{wang2025metis} & G                   & 3.588 & 3.980 & 3.265 & 3.882 & 3.869 & 0.879 & 0.180 & 0.890 \\
      \hline
      GenTSE               & G                         & \textbf{3.656} & \textbf{4.135} & \textbf{3.399} & \textbf{4.296} & \textbf{3.976} & \textbf{0.928} & 0.177 & \textbf{0.920} \\
      \Xhline{2\arrayrulewidth}
    \end{tabular}
    } 
    \label{tbl:compare_with_other_baselines}
  \end{center}
  \vspace{-10pt}
\end{table*}

\subsection{Frozen-LM Conditioning}
\label{subsec: Frozen-LM Conditioning}
 Exposure bias \cite{schmidt2019generalization, liu2021scheduled} arises from the mismatch between ground-truth conditioning during training and self-generated histories at inference. To address this issue, we adopt a Frozen-LM Conditioning (FLC) strategy inspired by scheduled sampling \cite{mihaylova2019scheduled}. We first train the semantic and acoustic LMs using teacher forcing with Eqs. (\ref{eq:ce_loss_sem}) and (\ref{eq:ce_loss_aco}), obtaining base parameters $\theta$ and $\phi$. We then duplicate these parameters to form trainable models $\theta'$ and $\phi'$, while keeping $\theta$ and $\phi$ frozen. During FLC, the frozen models generate predicted semantic and acoustic tokens $\hat{S}$ and $\hat{A}$ under teacher forcing. These predictions are used as conditioning inputs when training $\theta'$ and $\phi'$, replacing the corresponding ground-truth conditioning tokens $s_{<t}$ and $a_{<n}$. This design exposes the target models to model-generated conditioning signals while avoiding instability caused by jointly updating the conditioning models. The resulting training objectives are defined below, where $\hat{s}'_t$ and $\hat{a}'_n$ are the generated semantic and acoustic tokens by $\theta'$ and $\phi'$ at timestamp $t$ and $n$: 

\begin{equation}
\mathcal{L}_{\text{CE}'_{\text{sem}}}
= - \sum_{t=1}^{T}
\log P_{\theta'}(
    \hat{s}'_t \mid \hat{s}_{<t}, Wavlm_{ref}, Wavlm_{mix}).
\label{eq:flc_sem}
\end{equation}

\begin{equation}
\mathcal{L}_{\text{CE}'_{\text{aco}}}
= - \sum_{n=1}^{N}
\log P_{\phi'}(
    \hat{a}'_n \mid \hat{a}_{<n}, S, DAC_{ref}, DAC_{mix}).
\label{eq:flc_aco}
\end{equation}

\subsection{Direct Preference Optimization}
\label{subsec: Direct Preference Optimization}
Existing LM-based TSE methods rely on token-level likelihood objectives, which do not strictly align with perceptual speech quality. To address this, we train the acoustic LM with DPO \cite{rafailov2023direct} to explicitly bias its output distribution toward perceptually preferred speech. For each context $y = [S, DAC_{ref}, DAC_{mix}]$, we construct a pair of Simcodec token sequences $(A^+, A^-)$, where $A^\pm = [a^\pm_1, \ldots, a^\pm_N]$ and $A^+$ is judged better than $A^-$. The DPO objective then optimizes the model to increase the likelihood of $A^+$ while correspondingly decreasing the likelihood of $A^-$, as specified in Eq. (\ref{eq:dpo_factored}).

\begin{equation}
\mathcal{L}_{\mathrm{DPO}}
= -\mathbb{E}\Bigg[
\log \sigma \Big(
\beta \log 
\frac{\pi_\psi({A^+\mid y})\, \pi_{\mathrm{ref}}({A^-\mid y})}
{\pi_{\mathrm{ref}}({A^+\mid y})\, \pi_\psi({A^-\mid y})}
\Big)
\Bigg].
\label{eq:dpo_factored}
\end{equation}

Here, $\psi$ denotes the parameters of the target model being optimized and $\mathrm{ref}$ represents the parameters of a frozen reference model used to stabilize training. Both models are initialized to $\phi'$, the parameters after FLC. $\pi$ is the sequence probability distribution, the scalar $\beta>0$ regulates preference strength, and $\sigma(\cdot)$ corresponds to the logistic sigmoid function.

To construct a preference pair $(A^+, A^-)$, the $ref$ model conditions on context $y$ to produce logits $Logits_{ref}(\hat{A}) \in \mathbb{R}^{N \times |\mathcal{V}|}$, where $|\mathcal{V}|$ is the vocabulary size. From these logits, $M$ candidate Simcodec token sequences are independently sampled via top-$k$ multinomial sampling at each timestep. Each candidate sequence is then decoded by the SimCodec decoder, and evaluated using UTMOS \cite{saeki2022utmos}, a DNN-based MOS predictor strongly correlated with human judgments. The candidate sequence with the highest UTMOS score is selected as $A^+$, while the one with the lowest score becomes $A^-$. 


\section{Experiments}
\label{sec:Experiments}

\subsection{Dataset}
We train all LMs using the train-100 and train-360 splits of the clean LibriMix2spk dataset \cite{cosentino2020librimix}, a standard benchmark for TSE, and conduct evaluation on the corresponding test set. For inference, a reference utterance of the target speaker is randomly chosen and used consistently in all evaluations. WavLM (WavLM-Large), DAC\footnote{\href{https://huggingface.co/hance-ai/descript-audio-codec-16khz}{https://huggingface.co/hance-ai/descript-audio-codec-16khz}} and SimCodec\footnote{\href{https://huggingface.co/yaoxunji/gen-se}{https://huggingface.co/yaoxunji/gen-se}} are obtained using their publicly available pretrained checkpoints. KMeans (for discretizing WavLM embedding) is trained on 960 hours of LibriTTS. All data is processed at a 16 kHz sampling rate.

\begin{table}[t]
  \caption{
Ablation on LM input conditions. ref, mix, tar denote conditions from reference, mixture, and target speech. W, D, S indicate WavLM, DAC, SimCodec; E and T denote continuous embeddings and discretized tokens. Exp. 5 is the final setup.
    }
  \centering
  \renewcommand{\arraystretch}{1.15}
  \setlength{\tabcolsep}{5pt}

  \resizebox{\columnwidth}{!}{
  \begin{tabular}{lcccccccc}
    \Xhline{2\arrayrulewidth}
    Exp & \multicolumn{2}{c}{Semantic LM} & \multicolumn{3}{c}{Acoustic LM} & UTMOS $\uparrow$ & dWER $\downarrow$ & SECS $\uparrow$ \\
    \cline{2-6}
        & ref & mix & ref & mix & tar &  &  &  \\
    \hline
    1 & -   & -   & D-E & D-E & - & 4.007 & 0.284 & \textbf{0.928} \\
    2 & W-T & W-T & D-E & D-E & W-T & 3.734 & 0.738 & 0.839 \\   
    3 & W-E & W-E & S-T & S-T & W-T & \textbf{4.112} & 0.237 & 0.906 \\
    4 & W-E & W-E & W-E & D-E & W-T & 4.080 & 0.220 & 0.927 \\
    \hline
    5 & W-E & W-E & D-E & D-E & W-T & 4.065 & \textbf{0.217} & 0.927 \\
    \Xhline{2\arrayrulewidth}
  \end{tabular}
  }
  \label{tbl:exp_on_input_conditions}
\end{table}

\begin{table}[t]
  \caption{FLC and TF fine-tuning. Initialized from Exp. 5.}
  \centering
  \renewcommand{\arraystretch}{1.15}
  \setlength{\tabcolsep}{5pt}

  \resizebox{\columnwidth}{!}{
  \begin{tabular}{lcccccc}
    \Xhline{2\arrayrulewidth}
    Exp & \multicolumn{1}{c}{Method} & \multicolumn{1}{c}{Sem/Aco steps} & UTMOS $\uparrow$ & dWER $\downarrow$ & SECS $\uparrow$ \\
    \hline
    6 & TF & 12k / 6k & 4.118 & 0.189 & 0.933 \\   
    7 & TF & 20k / 20k & 4.121 & 0.184 & 0.933 \\  
    \hline
    8 & FLC & 12k / 6k & \textbf{4.127} & \textbf{0.172} & \textbf{0.935} \\  
    \Xhline{2\arrayrulewidth}
  \end{tabular}
  }
  \label{tbl:exp_on_flc}
\end{table}

\begin{figure}[!t]
    \includegraphics[scale=0.35]{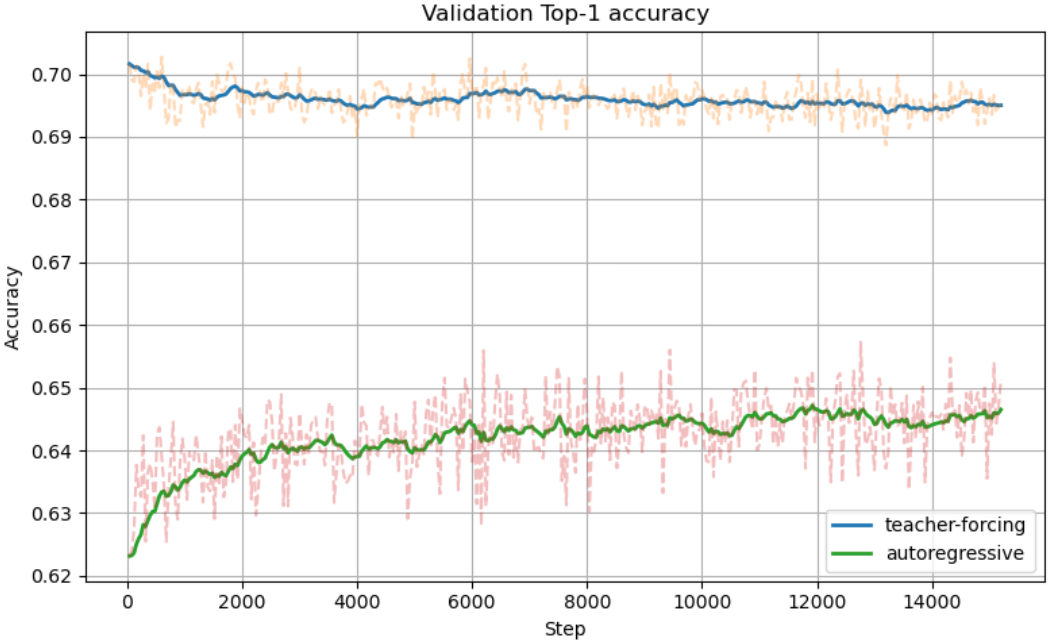}
    \caption{Semantic LM validation top-1 accuracy during FLC.}
    \label{fig:semantic_lm_FLC_curve} 
\end{figure}

\begin{table*}[t]
  \caption{DPO and CE fine tuning. Initialized from Exp. 8. Bold/underline indicate best/second best results.}
  \renewcommand{\arraystretch}{1}
  \begin{center}
    \resizebox{\textwidth}{!}{
    \begin{tabular}{ccccccccccc}
      \Xhline{2\arrayrulewidth}
      \multirow{2}{*}{Exp} & \multirow{2}{*}{Method} & \multicolumn{3}{c}{DNSMOS $\uparrow$} & \multirow{2}{*}{UTMOS $\uparrow$} & \multirow{2}{*}{NISQA $\uparrow$} & \multirow{2}{*}{SECS $\uparrow$} & \multirow{2}{*}{dWER $\downarrow$} & \multirow{2}{*}{SpeechBERT $\uparrow$} \\
      \cline{3-5}
                             &                           & SIG & BAK & OVL &                         &                      &                         &                          &                          \\
      \hline
      8 & - & 3.635 & 4.111 & 3.366 & 4.127 & 3.806 & \textbf{0.935} & \textbf{0.172} & \underline{0.919} \\
      \hline
      9 & $\mathcal{L}_{\mathrm{CE}}$                   & 3.633 & 4.102 & 3.359 & 4.125 & 3.824 & \underline{0.934} & \underline{0.173} & \textbf{0.920} \\
      10 & $\mathcal{L}_{\mathrm{DPO}}$                   & \textbf{3.672} & \textbf{4.161} & \textbf{3.429} & \textbf{4.384} & \textbf{4.108} & 0.913 & 0.182 & 0.915 \\
      11 & $\mathcal{L}_{\mathrm{DPO}} + \mathcal{L}_{\text{CE}}$           & \underline{3.656}
 & \underline{4.135} & \underline{3.399} & \underline{4.296} & \underline{3.976} & 0.928 & 0.177 & \textbf{0.920} \\
      \Xhline{2\arrayrulewidth}
    \end{tabular}
    } 
    \label{tbl:exp_on_dpo}
  \end{center}
  \vspace{-10pt}
\end{table*}

\subsection{Evaluation Metrics}
We evaluate speech quality using 3 DNN-based metrics: DNSMOS \cite{reddy2021dnsmos}, UTMOS \cite{saeki2022utmos} and NISQA \cite{mittag2021nisqa}. DNSMOS provides three sub-scores, SIG for signal quality, BAK for background noise, and OVRL for overall quality. UTMOS and NISQA are also speech quality predictors and align strongly with human judgments. For NISQA, we use the NISQA-TTS (v1.0) checkpoint. Speech intelligibility is assessed using the Differential Word Error Rate (dWER) and SpeechBERT \cite{saeki2024speechbertscore}. dWER is computed by comparing the Whisper-base \cite{radford2023robust} transcriptions of the predicted and target speech. SpeechBERT computes BERTScore as a semantic-similarity metric, using SSL features derived from the generated and target speech signals. We employ the HuBERT-base \cite{hsu2021hubert} model for feature extraction. Target-speaker preservation is measured using speaker embedding cosine similarity (SECS), calculated using Resemblyzer\footnote{\href{https://github.com/resemble-ai/Resemblyzer}{https://github.com/resemble-ai/Resemblyzer}}. We do not report classical metrics such as PESQ or SI-SNR, as generative models may introduce temporal or prosody misalignments that render these metrics unreliable \cite{wang2024selm, tang2024tselm, yao2025gense, zhang2025anyenhance}.

\subsection{Implementation Details}
The semantic and acoustic language models are decoder-only Transformers with 12 layers, 8 attention heads, and a hidden size of 1024. The K-means and SimCodec models use a cluster and codebook size of 1024 and 8192. All experiments use the AdamW optimizer. Either A40 or A100 GPUs are used.

For all ablation on the choice of the LMs' input conditions (Exps 1-5), we train the semantic LM $\theta$ on 1 GPU with batch size 32, and the acoustic LM $\phi$ on 4 GPUs with batch size 64. Both use a peak learning rate of $1\times10^{-4}$ with 1k warmup steps. Trainings are terminated upon validation cross-entropy convergence. The final configuration (Fig. \ref{fig:overall_pipeline}, Exp 5) corresponds to 64k steps for $\theta$ and 78k steps for $\phi$.

For FLC (Exp 8) and comparative TF (Exp 6), $\theta'$ and $\phi'$ are trained for 12k and 6k steps with batch sizes 32 and 64, respectively, using a constant learning rate of $5\times10^{-6}$ on a single GPU, initialized from $\theta$ and $\phi$ in Exp 5. In Exp 6, the validation cross-entropy continues to decrease under this updated learning rate. Accordingly, in Exp. 7 we repeat the Exp 6 setting but extend training of $\theta'$ and $\phi'$ to 20k steps, until convergence.

For DPO experiments (Exp 9 to 11), training is initialized from $\phi'$ in Exp 8 and performed on 1 GPU for 400 steps with batch size 128 and learning rate $5\times10^{-6}$. We set $\beta{=}0.1$ (Eq.~\ref{eq:dpo_factored}), $k{=}16$ for top-$k$ sampling, and generate $M{=}32$ candidates.

\subsection{Baseline Models}
We compare GenTSE against recent generative LM-based baselines (TSELM-L \cite{tang2024tselm}, LLaSE-G1 \cite{kang2025llase}, Metis \cite{wang2025metis}) as well as discriminative baselines (X-TF-GridNet \cite{hao2024x}, USEF-SepFormer \cite{zeng2025usef}). TSELM-L and USEF-SepFormer are retrained on our training set for a fair comparison. For LLaSE-G1, a multi-task model, we use the official checkpoint reported in the paper. For Metis, we adopt the released checkpoint fine-tuned for TSE \footnote{\href{https://huggingface.co/amphion/Metis}{https://huggingface.co/amphion/Metis}}. Note that LLaSE-G1 and Metis are both trained on substantially larger datasets. For X-TF-GridNet, we use the author-provided checkpoint trained on WHAMR! \cite{maciejewski2020whamr}.


\section{Results and discussion}
Table \ref{tbl:compare_with_other_baselines} compares GenTSE against recent generative LM-based and discriminative TSE models. GenTSE surpasses prior LM-based baselines in speech quality, speaker similarity, and intelligibility, demonstrating clear advantage among LM-driven methods. When compared with leading discriminative methods, GenTSE achieves higher perceptual speech quality and speaker similarity; however, USEF-SepFormer attains lower dWER.

Table \ref{tbl:exp_on_input_conditions} evaluates LM performance under different input conditions. Exp 5 corresponds to our final configuration; Exps 1-4 are ablations. In Exp 1, removing the semantic stage leads to increased dWER (Exp1: 0.284 vs. Exp 5: 0.217), highlighting the necessity of the semantic extraction stage. Exp 2 shows that discretizing WavLM embeddings in the semantic stage introduces substantial information loss, hindering accurate recovery of target-speaker semantics. Exp 3 demonstrates that conditioning the Acoustic LM on SimCodec tokens reduces intelligibility and speaker similarity compared to using DAC embeddings (Exp 5). Exp 4 shows that conditioning the Acoustic LM on WavLM embeddings (layer 6) of the reference speech yields comparable performance to using DAC embeddings. Collectively, these results validate our LM input design. 

Table \ref{tbl:exp_on_flc} compares TF and FLC. FLC (Exp 8) outperforms TF fine tuning (Exp 6) in same training steps, with notable lower dWER. Even when TF training is extended until no further improvement is observed (Exp 7), FLC (Exp 8) remains superior. Fig. 2 further shows the gap in the top-1 validation accuracy of the Semantic LM, which progressively narrows under FLC, demonstrating its effectiveness in mitigating exposure bias. 

In table \ref{tbl:exp_on_dpo}, Exps 9 to 11 apply either CE (Eq. \ref{eq:ce_loss_aco}),  DPO (Eq. \ref{eq:dpo_factored}) or the sum of both loss. No loss scaling is used in Exp 11 due to comparable magnitudes. Exp 10 and 11 yield consistent improvements in speech-quality metrics (DNSMOS, UTMOS, NISQA) after only 400 steps, indicating the effectiveness of DPO in enhancing perceptual speech quality. Minor drops in speaker similarity and intelligibility reflect a trade-off, as the current DPO setup prioritizes perceptual quality. Incorporating multi-metric preference selection \cite{zhang2025multi} is a promising direction for future work. Using only DPO loss (Exp 10) provides more quality gain, but also additional reductions in similarity and intelligibility. In contrast, fine tuning with cross-entropy loss alone (Exp 9) yields small difference from Exp 8.


\section{Conclusion}
We present GenTSE, a fully generative TSE framework based on a two-stage decoder-only LM architecture. The semantic stage predicts k-means-discretized WavLM tokens of the target speech, and the acoustic stage generates neural audio codec tokens conditioned on these predictions. This two-stage design reduces modeling complexity compared to direct codec-token prediction, improving performance. Both LMs are conditioned on continuous frame-level WavLM/DAC embeddings from the reference and mixture speech, providing richer guidance than discrete input. To address the mismatch between teacher-forcing and autoregressive inference, we introduce FLC which fine-tunes the LMs on predicted tokens from early frozen checkpoints. Additionally, we use DPO to align generation with perceptual preferences via proxy-MOS feedback. Extensive experiments on Libri2Mix validate the architecture and optimization strategies, showing superior performance over prior LM-based methods. Future research may further investigate the robustness of GenTSE under noisy and target-absent conditions.

\newpage
\clearpage

\section{Acknowledgments}
This research is supported by the RIE2025 Industry Alignment Fund - Industry Collaboration Projects (IAF-ICP) (Grant No. I2501E0041), administered by A*STAR, as well as supported by Schaeffler (Singapore) PTE. LTD. and NTU Singapore through Schaeffler-NTU Corporate Lab: Intelligent Mechatronics Hub.

\section{Generative AI Use Disclosure}
Generative AI tools were used for limited editorial support, including grammar checking, redundant text removal, and assistance with LaTeX equations. All scientific work, such as the introduction, methodology, experiments, results, and conclusions, was conducted by the authors. All authors reviewed the manuscript and take full responsibility for the final submission.

\bibliographystyle{IEEEtran}
\bibliography{mybib}

\end{document}